\documentstyle[12pt,epsf]{article}
\newcommand {\be}{\begin{equation}}
\newcommand {\ee}{\end{equation}}

\begin{document}
\title{Mean Field Behaviour in a Local Low-Dimensional Model}
\author{Hans-Martin Br\"oker and Peter Grassberger\\
Physics Department, University of Wuppertal\\
42097 Wuppertal, Germany}
\maketitle

\begin{abstract}
We point out a new mechanism which can lead to mean field type behaviour
in nonequilibrium critical phenomena. We demonstrate this mechanism on
a two-dimensional model which can be understood as a stochastic and
non-conservative version of the abelian sandpile model of
Bak {\it et al.} ~\cite{bak}. This model has a second order phase
transition whose critical behaviour seems at least partly described
by the mean field approximation for percolation, in spite of the low
dimension and the fact that all interactions are of short range.
Furthermore, the approximation obtained by replacing the lattice by a
Bethe tree is very precise in the entire range of the control parameter.
\end{abstract}
\newpage

Although mean field theory is one of the most useful and most commonly
applied approximations for spatially extended complex systems, it has
an important drawback: near critical
points it is usually not only quantitatively but also qualitatively
wrong, in the sense that it predicts wrong critical exponents and
scaling functions. There are only few known exceptions to this. They
involve systems with infinitely long range interactions, systems in
high dimensions, and systems defined on ``lattices" without loops
(``Bethe tree" lattices).

The reason why mean field theory gives wrong results is easily understood
qualitatively. By neglecting correlations, it also neglects the feedback
of the local order parameter value ${s(\mathbf x})$ onto itself. Since a
system at its critical point is very sensitive to any influence, these
feedbacks are enhanced and modify its behaviour in an essential way. The
above exceptions avoid this problem by either disallowing feedback loops
(Bethe lattices) or by suppressing the effect of short loops in favour of
long loops which cover essentially the whole lattice.

In this note we want to point out a novel mechanism for generating mean
field behaviour. This mechanism can only be effective in non-equilibrium
systems, but we conjecture that it should occur there rather commonly.
In equilibrium systems, the above-mentioned feedback is always
{\it positive} due to detailed balance. In non-equilibrium systems,
however, the feedback can also be negative. If the system reacts to the
feedback in a strongly non-linear way, the effects of individual
loops will thus not be additive, but might cancel. In general, this
cancellation will of course not be complete, and the net effect of the
feedback will be similar to random noise. Otherwise said, the
order parameter $s({\mathbf x})$ will be ``confused" by the influences of
the different feedback loops. This confusion will lead to a fast
loss of memory which in turn implies that the mean field approximation
can become correct.

Specifically, we shall illustrate this with a model which is inspired
by the sandpile model \cite{bak}. It is also abelian in the sense
of \cite{dhar}, but the evolution of avalanches is stochastic and
non-conservative. As a consequence, its critical behaviour is not
{\it self-organized}, but it has a critical point in the ordinary
(co-dimension 1) sense. A model which gives identical spatial
structures was first introduced by
one of us in \cite{grass} (model nr. 5 in the appendix of that paper).
Similar (but somewhat more complicated) models were also studied
in ~\cite{man}. Although we cannot prove analytically that the critical
behaviour of our model is of mean field type, our numerical results
suggest very strongly that at least two of the three independent critical
exponents have their mean field value, and that scaling functions are
identical.

Our model is defined on a two-dimensional square lattice (generalizations
to higher dimensions are obvious), and time is discrete. At each lattice
site we have a ``spin" $z_{i}$, which can take any non-negative integer
value, but only the values $z_i=0$ and $z_i=1$ are "stable". If $z_i$
becomes $>1$ during the evolution, it ``topples" at the next time step.
The toppling rule is
\be
    z_{i}\; \to \; z_{i}-2                       \label{tpe}
\ee
for the site which topples, and
\be
   z_{j} \to \left\{ \begin{array}{l@{\quad:\quad}l}
          z_{j}+1 & \mbox{with probability}\quad p\\
          z_{j}   & \mbox{with probability}\quad 1-p
                  \end{array}    \right.           \label{tpz}
\ee
for each of its four neighbours. Notice that each neighbour $j$ has the
same chance $p$ to get its spin increased, {\it independently} of what
happens at the other neighbours. Thus the sum $\sum_iz_i$
fluctuates (during each toppling, it can change by any value between
-2 and +2), but in the average each toppling event causes
$\sum_iz_i$ to decrease by $4p-2$. The critical
point is exactly where this vanishes, $p_c = 1/2$. For later use we
define $\epsilon=1/2-p$, and $\varrho = \langle z \rangle$.

As in the sandpile model, the dynamics consists of two rules: \\
(i) If all sites are stable, a site $i$ is chosen randomly, and $z_i$ as
well as time $t$ are increased by one unit. In the following, we call
this an ``event".\\
(ii) If at least one site is unstable, the above toppling rule is applied
simultaneously at all unstable sites. If some $z_i\geq 4$ so that it has
to topple repeatedly to become stable, then all these topplings are also
done simultaneously. After this, $t$ is increased by 1. This is repeated
until all sites are stable again, after which rule (i) is applied again.

Our most extensive studies were done for $p<p_c$, where all avalanches
are finite (with probability one), even on an infinite lattice. In this
case we can use periodic boundary conditions, which greatly reduces
finite-size effects. Some simulations were done also at $p=p_c$. There
$\varrho$ would not decrease during an avalanche,
leading to problems with infinite avalanches when periodic boundary
conditions are used \footnote{Serious problems arise only when
$p>p_c$, but we expect very slow behaviour and numerical instabilities
already at the critical point.}. In this case, we used the same kind of
open b.c. as in the sandpile model: if a neighbour of site $i$ is outside
the boundary, we just disregard it in eq.(\ref{tpz}).

In a mean field theory for this model, we assume that a toppling site
"forgets" after the next time step that it has toppled, and $z_{i}$ is
replaced by an identically and independently distributed random variable
which takes the value 1 with probability $\varrho$, and 0 with probability
$1-\varrho$. This implies in particular that there are no correlations,
$c(i-j)\equiv\langle z_iz_j\rangle-\varrho^2=0$. Correlations measured in
simulations are shown in fig.1. We see that they are indeed small and,
what is more important, they decay very rapidly with distance. A similarly
rapid decay was observed also in the sandpile model \cite{gr-man}. In
that model it was indeed proven that the correlations decay as
$||i-j||^{-4}$ \cite{dhar2}.

\begin{figure}[htp]
%\centerline{\epsfbox{corr.ps}}
\caption{\small Correlations measured in simulations, plotted against the
square of the Euclidean distance. For distances $>4$, all correlations
were compatible with zero.}       \label{cc.fig}
\end{figure}

Notice that we assume that this loss of memory
due to confusion by the topplings of the other neighbours happens only
{\it after} the next time step. So each toppling site (except the very
first of the avalanche which we call the {\it root}) has three
neighbours which it can cause to become unstable with probability
$a=\varrho p$, while the fourth neighbour (its father) will be induced
to topple with a different probability $a'$. The simplest
assumption would be $a'=0$. More precise estimates can be obtained from
self-consistency arguments or from comparison with simulations. Both
give $a'/a$ \raisebox{-2pt}{$\stackrel{<}{\sim}$} $0.1$. In the following we
shall compare simulation
data with predictions for the two extreme choices $a'=0$ and $a'=0.1a$,
mainly to show that the results depend very little on $a'$.

For $a'=0$ our mean field treatment would be completely equivalent to
percolation on a Bethe lattice with coordination number four, the sites
of which are occupied with probability $a$ (except for the root, which
is occupied with probability $\varrho$). This problem is exactly
soluble and well discussed in the literature \cite{stauffer}.
The extension to $a'\neq 0$ is a straightforward application of the
theory of branching processes \cite{har}.

But first we have to compute $\varrho$ and $a$ as functions of the
controll parameter $p$. For this we use stationarity and the fact that
each toppling decreases $\langle \sum_iz_i\rangle$ by $4\epsilon$, while
each event (whether it actually triggers off an avalanche or not)
increases it by one unit. Thus the average number of topplings per event
is $\langle s(p) \rangle =1/4\epsilon$. On the other hand, the average
number of topplings during the first update of an avalanche is
$4a$, while it is multiplied by $3a+a'$ during each successive update.
Taking into account that each event triggers an avalanche with probability
$\varrho$, we find
\be
   \langle s(p)\rangle=\varrho[1+4a\sum_{i=0}^{\infty}
     (3a+a')^i] = \varrho \frac{1+a-a'}{1-3a-a'}\,.        \label{sp}
\ee
Combining these two estimates, we find a somewhat complicated expression
for $\varrho$ as a function of $p$. Instead of writing it down we just
give inversely $\varrho$ and $p$ as functions of $a$ and $a'$:
\be
    \varrho={1\over2}+2a{a-a'\over 1+a-a'}, \quad p=a/\varrho\;. \label{rp}
\ee
This gives always $p\leq 1/2$, as we should expect since no stationary
solution exists for $p> 1/2$. For $p\to 1/2_-$, we find $\varrho =
{2\over3+(a'/a)}[1-\epsilon{2(1-a'/a)(7+a'/a)\over(3+a'/a)^2}]$.
Fig.2 shows the very reasonable agreement of this prediction
with values obtained by simulations. We point out in particular that
the data show that $d\varrho/dp$ is finite at $p\to 1/2$,
$\varrho\approx 0.6483-0.73\epsilon$, as predicted by eq.(\ref{rp}).

\begin{figure}[htp]
%\centerline{\epsfbox{rho.ps}}
\caption{\small Density \protect{$\varrho(p)$} against $p$. The
continuous line is the mean field prediction of
eq.(\protect{\ref{rp}}) with $a'=0$, the
dashed line is the prediction with $a'=0.1a$, and the points
show the numerical data obtained from simulations.}    \label{rho.fig}
\end{figure}

The quantity easiest to compute is the survival probability $P_t$,
defined as the probability that an event triggered an avalanche which
lasts for $\geq t$ time steps. We denote by $Q_t$ the probability that
all sites are again stable at time $t$, provided that the event started
with an unstable root at time 0. Then obviously $P_t=\varrho(1-Q_t)$.
Similarly, we call $\tilde{Q}_t$ the probability that an unstable site
different from the root will not create any unstable
offspring $t$ time steps later.

Furthermore, we denote by $p_k={4\choose k}(1-a)^{4-k}a^k$ the probability
that the first toppling triggers off $k$ topplings at the next time
step, and by $\tilde{p}_k$ the analog distribution for a later toppling.
Their generating functions are given by
\be
    g(s) = \sum_{k=0}^4 s^kp_k = (1-a+sa)^4
\ee
and
\be
    \tilde{g}(s) = \sum_{k=0}^4 s^k\tilde{p}_k = (1-a'+sa')(1-a+sa)^3 \;.
\ee
Then we have obviously
\be
   \tilde{Q}_t = \sum_k \tilde{p}_k [\tilde{Q}_{t-1}]^k =
            \tilde{g}(\tilde{Q}_{t-1})                 \label{tildQ}
\ee
and
\be
   P_t = \varrho(1-Q_t) = \varrho g(\tilde{Q}_{t-1})\;.  \label{PQ}
\ee
In the vicinity of the critical point, $P_t$ obeys the scaling law
\cite{stauffer}
\be
    P_t \approx {1\over t^\delta} \Phi(t\epsilon^{\nu_t})  \label{pt.eq}
\ee
with $\delta = \nu_t =1$.

\begin{figure}[htp]
%\centerline{\epsfbox{dura.ps}}
\caption{\small Avalanche survival probabilities $P_t$ for different values of
$\epsilon$. From left to right $\epsilon=0.1,0.0316,0.01,0.00316,0.001$,
and 0.000316. Solid lines are from mean field theory with $a'=0$, dashed
lines from mean field theory with $a'/a=0.1$, and dots are from
simulations.}                  \label{pt.fig}
\end{figure}

Predictions for $P_t$ are compared to simulation data in
fig.\ref{pt.fig}. Notice that we used there the exact predictions
from the recursion relation eq.(\ref{tildQ}) which can be made
numerically stable by some minor rearrangements. For small values of
$t$ we see perfect agreement, while there are deviations at very
large $t$. They indicate that $\nu_t$ is somewhat larger than its mean
field value, $\nu_t = 1.024\pm 0.008$, while $\delta=1.00\pm 0.01$ is
in exact agreement with the prediction.

The other quantity we studied is the avalanche size distribution
${\cal P}_n$, defined as the probability that an event involves exactly
$n$ topplings. For $n=0$ and $n=1$ we have ${\cal P}_0=1-\varrho$ and
${\cal P}_1=\varrho(1-a)^4$. To calculate ${\cal P}_n$ for $n>1$ we use
a theorem due to Dwass\cite{dwass}. Consider a branching process where
the distribution of offsprings of a single individuum is given by $p_k$,
with generating function $g(s)$. Its {\it progeny} is the total number
of descendants, including itself. Then the probability that the total
progeny of $n_0$ individua is $n$ is given by \cite{dwass}
\be
   P_{n|n_0} = {n_0\over n} p_{n-n_0}^{(n)}\;, \quad n\geq n_0\;,
\ee
where $p_k^{(n)}$ is the $k$-th Taylor coefficient of $[g(s)]^n$, i.e.
$[g(s)]^n = p_0^{(n)}+p_1^{(n)}s+p_2^{(n)}s^2+\ldots$.

Denoting again quantities referring to side branches by tilde's, we
have in the present case
\be
    {\cal P}_n = \varrho\sum_{k=0}^4 p_k \tilde{P}_{n-1|k} =
      \varrho\sum_{k=0}^4 {k\over n-1} p_k \tilde{p}_{n-k-1}^{(n-1)}\;,
\ee
while $\tilde{p}_k^{(n)}$ can be computed in several ways. The most
straightforward is to use binomial expansions in $[\tilde{g}(s)]^n$.
For large values of $n$ we can also use the asymptotic behaviour.
{}From the cumulant expansion of $\tilde{g}(s)$ we see that
$\tilde{p}^{(n)}$ tends for $n\to\infty$ towards a Gaussian probability
distribution with mean value $n\langle k\rangle$ and variance $n\sigma$,
where $\langle k\rangle=3a+a'$ and $\sigma=3a(1-a)+a'(1-a')$ are the
first two cumulants of $\tilde{p}$. This gives us
\be
    \tilde{P}_{n|n_0} \approx {n_0\over \sqrt{2\pi n^3\sigma}}
      e^{-{[(1-\langle k\rangle)n-n_0]^2\over2n\sigma}} \;. \label{asy}
\ee

\begin{figure}[htp]
%\centerline{\epsfbox{ener.ps}}
\caption{\small Integrated avalanche size distribution \protect{$D_n$} for
the same values of $\epsilon$ as in fig.3. Solid lines are again
predictions with $a=0$, dashed lines are from predictions with
$a'=0.1a$, and points are from simulations.}     \label{pn.fig}
\end{figure}

The integrated distribution $D_n=\sum_{m=n}^\infty {\cal P}_m$ is shown
in fig.4. Again, the continuous and dashed lines give the predictions
with $a'=0$ and $a'=0.1a$, while the dots are results from simulations.
This time the agreement is essentially perfect. This proves also that the
simulation data satisfy the mean field scaling behaviour obtained from
eq.(\ref{asy}),
\be
    D_n = {1\over \sqrt{n}} \Psi(n\epsilon^2) \;.
\ee
But again we found that the agreement with the detailed prediction of the
Bethe lattice model is much better than that with the scaling form.

Finally, we mention that we also performed some simulations at $p=1/2$,
on lattices of size $L\times L$ with open boundary conditions. We found
that the average avalanche size $\langle n\rangle$ scales as $L^2$, as
expected for any model of this type in which $\langle\sum_iz_i\rangle$ is
conserved \cite{dhar,kadano}.

In summary, we have shown numerically that a non-equilibrium model inspired
by sandpile models shows a critical behaviour which is extremely close to
mean field type. Indeed it seems that all exponents except one are identical
to those for mean field percolation. This is so in spite of the fact that
the model involves only short range interactions and lives on a 2-$d$
lattice. We have argued that this is due to the ``confusion" brought about
by the non-positive and non-linear effect of feedback loops.

We should mention that similar
behaviour was observed earlier in similar models \cite{man}. But
apart from being more complicated, these models were still closer in
spirit to the original sandpile model of \cite{bak}. In particular,
in \cite{man} models were studied were the avalanche evolution was either
stochastic or non-conservative, but not both together. It was the
combination of both aspects which allowed us to draw a very close
connection to percolation, and to compare in detail with percolation
on a Bethe lattice.

Finally, we should mention that our model is very
similar -- superficially seen -- to the Ising model. Indeed, only
very minor modifications were needed in the computer program given
in \cite{grass} to switch from one model to the other. The main difference
with the Ising model is that the present model does not satisfy detailed
balance and is thus an inherently non-equilibrium system. We believe that
the latter is necessary to observe mean-field type behaviour generated
by the above mechanism.

\vspace{.5cm}

P.G. wants to thank L. Pietronero for interesting discussions, and A.
Vulpiani for hospitality at the University of Rome where part of this work
was done. It was also supported by the DFG (Sonderforschungsbereich 237).

\end{document}